\documentclass[twocolumn,prb,showpacs,superscriptaddress,preprintnumbers,amsmath,amssymb]{revtex4}

\usepackage{graphicx}
\usepackage{dcolumn}
\usepackage{bm}
\begin{document}
\title{Low energy excitations in CoO studied by temperature dependent x-ray absorption spectroscopy}

\author{M. W. Haverkort}
  \affiliation{II. Physikalisches Institut, Universit{\"a}t zu K{\"o}ln,
   Z{\"u}lpicher Str. 77, D-50937 K{\"o}ln, Germany}
  \affiliation{Max Planck Institute for Solid State Research, Heisenbergstra{\ss}e 1, D-70569 Stuttgart, Germany}
\author{A. Tanaka}
  \affiliation{Department of Quantum Matter, ADSM, Hiroshima University, Higashi-Hiroshima 739-8530, Japan}
\author{Z. Hu}
  \affiliation{II. Physikalisches Institut, Universit{\"a}t zu K{\"o}ln,
   Z{\"u}lpicher Str. 77, D-50937 K{\"o}ln, Germany}
\author{H. H. Hsieh}
  \affiliation{Chung Cheng Institute of Technology, National Defense University, Taoyuan 335, Taiwan}
\author{H.-J. Lin}
  \affiliation{National Synchrotron Radiation Research Center, 101 Hsin-Ann Road, Hsinchu 30077, Taiwan}
\author{C. T. Chen}
  \affiliation{National Synchrotron Radiation Research Center, 101 Hsin-Ann Road, Hsinchu 30077, Taiwan}
\author{L. H. Tjeng}
  \affiliation{II. Physikalisches Institut, Universit{\"a}t zu K{\"o}ln,
   Z{\"u}lpicher Str. 77, D-50937 K{\"o}ln, Germany}

\date{\today}

\begin{abstract}
We have measured the intricate temperature dependence of the Co
$L_{2,3}$ x-ray absorption spectra ($2p$-$3d$ excitations) of CoO.
To allow for accurate total electron yield measurements, the
material has been grown in thin film form on a metallic substrate
in order to avoid charging problems usually encountered during
electron spectroscopic studies on bulk CoO samples. The changes in
spectra due to temperature are in good agreement with detailed
ligand-field calculations indicating that these changes are
mostly due to thermal population of closely lying excited states,
originating from degenerate $t_{2g}$ levels lifted by the
spin-orbit coupling. Magnetic coupling in the ordered phase,
modeled as a mean-field exchange field, mixes in excited states
inducing a tetragonal charge density. The spin-orbit coupling
induced splitting of the low energy states results in a
non-trivial temperature dependence for the magnetic
susceptibility.
\end{abstract}

\pacs{71.70.Ej, 75.10.Dg, 78.70.Dm}
\maketitle

It is an observation that many insulating $3d$ transition metal
compounds do not reveal orbital degrees of freedom in their
electronic structure despite the incomplete filling of the $3d$
shell. This is related to the fact that the symmetry of the
crystal often becomes so low that the degeneracy of orbitals is
effectively lifted. This could be seen as a consequence of the
Jahn-Teller theorem stating that: "when the orbital state of an
ion is degenerate for symmetry reasons, the ligands will
experience forces distorting the nuclear framework until the ion
assumes a configuration both of lower symmetry and of lower
energy, thereby resolving the degeneracy". \cite{Jahn37, Jahn38}
Nevertheless, for some compounds, with CoO and FeO as well known
examples, a (nearly) cubic structure exists although these
systems have a partially filled $t_{2g}$ sub-shell. In such cases
the orbital degeneracy can be lifted by the spin-orbit coupling
(SOC). This in turn leads to quite interesting local physics, not
only because of the formation of orbital moments and large
magnetocrystalline anisotropies, but also because of a strong
temperature dependence in the local magnetic properties as we will address
below.

Taking the case of CoO, we will describe its ground state and
low-energy local excitations using the theoretical frame work laid
out by Kanamori \cite{Kanamori57} and Goodenough
\cite{Goodenough68}. The Co ions in this material are divalent
and have 7 electrons in the $3d$ shell. Due to electron-electron
interactions, a Hunds-rule $S=3/2$ high-spin ground-state is
realized, with an electronic configuration of approximately
$t_{2g}^5e_{g}^2$.\cite{Sugano70} The single hole in the $t_{2g}$
shell can be in three different orbitals, normally denoted as an
$xy$, $xz$, or $yz$ orbital. With the SOC being present, one
should consider making complex linear combinations of these
orbitals like $\sqrt{1/2}(-xz - \imath yz)$ that carry an orbital
momentum of 1 $\mu_B$ in the $z$ direction.\cite{Jo98} Orbital
momentum in the $x$ or $y$ direction can be created by cyclic
permutation of the coordinates. One can therefore assign to each
$t_{2g}$ electron a pseudo orbital momentum of $\tilde{l}=1$
(note that it is still a $d$ electron so
$l=2$.)\cite{Kanamori57,Goodenough68,AbragamBleaney} For CoO this
results in a total pseudo orbital momentum of $\tilde{L}=1$. This
orbital momentum couples with the spin $S=3/2$ to a state with
$\tilde{J}=1/2$, $3/2$, and $5/2$. The doublet with
$\tilde{J}=1/2$ is the ground-state, with the quartet about 40
meV higher and the sextet about 120 meV higher than the
ground-state. Furthermore, the sextet is split by the cubic
crystal field into a quartet and doublet. In estimating the
numbers we have used a SOC constant of $\zeta$ = 66 meV from
atomic Hartree-Fock calculations.\cite{HFValuesSO,Cowan81}

Experimentally one could measure these excitations with the use
of inelastic neutron scattering. The interpretation is somewhat
complicated as one measures phonons, magnons and spin-orbit
excitations all at the same time, but satisfying results have been
obtained.\cite{Sakurai68, Tomiyasu06} Yet, it would be welcome to
have an alternative spectroscopic method with perhaps also
advantages in terms of element specificity in case one is dealing
with a multicomponent materials, or higher material sensitivity for, e.g., thin films. In this regard it has been
recognized that soft x-ray absorption spectroscopy (XAS) should
have the potential to reveal the presence of those excited states:
Tanaka and Jo,\cite{Tanaka91} and De Groot \cite{deGroot94} have
calculated that theoretically one should see a strong temperature
dependence in the XAS lineshape. In such an XAS process, the
soft-x-ray photon (770-800 eV) is used to promote a $2p$ Co core
electron into the $3d$ valence shell. This transition is subject
to strict dipole selection rules, e.g. $\Delta J=0,\pm1$ in
spherical symmetry, with the result that each particular initial
state has its own set of reachable final states. It turned out
that the differences between the sets of final states are
appreciable when the differences in the initial states can be
traced back to differences in the quantum number $J$. In cubic
symmetry as for CoO, $J$ is not a good quantum number, but
following analogous arguments each initial state with a different
$\tilde{J}$ will have a different spectrum.

\begin{figure}
    \includegraphics[width=0.45\textwidth]{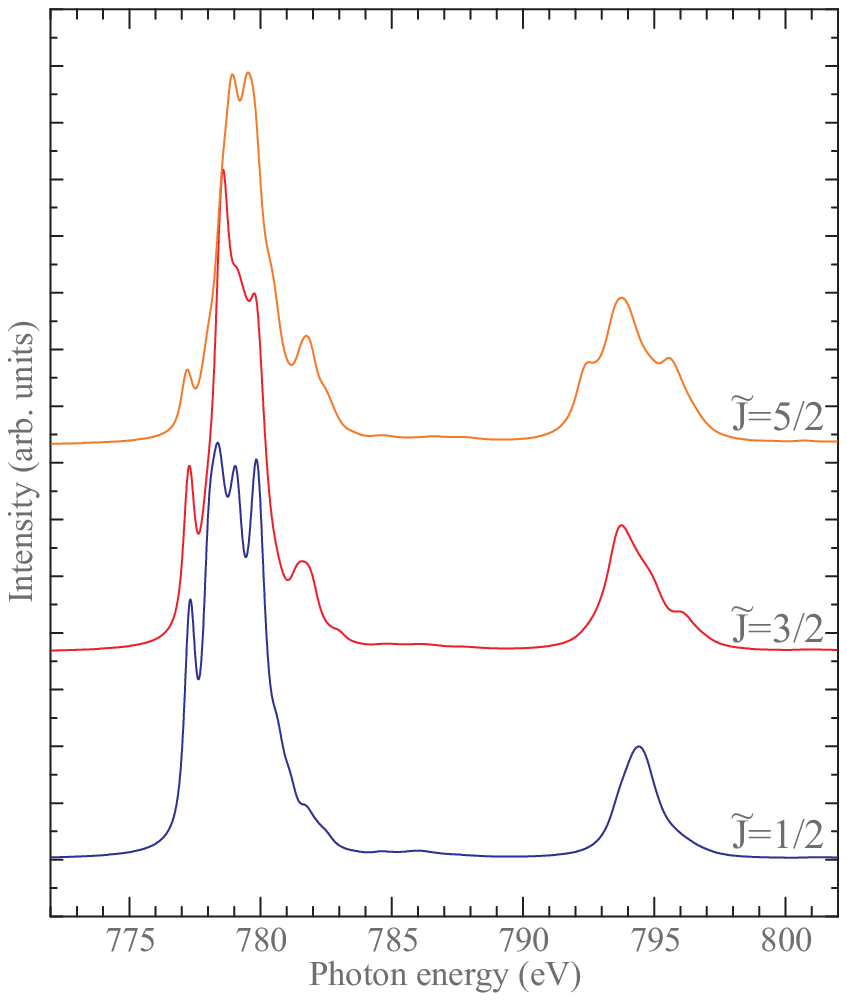}
    \caption{(color online) Theoretical isotropic Co-$L_{2,3}$ XAS spectra for
    the different initial states split by spin-orbit coupling as present in CoO.}
    \label{fig1}
\end{figure}

In Fig. 1 we have plotted the isotropic Co-$L_{2,3}$ XAS spectra
calculated for each initial state having a different $\tilde{J}$.
Here we used the successful configuration interaction model that
includes the full atomic multiplet theory and the hybridization
with the O $2p$ ligands.\cite{deGroot94,Tanaka94} We have carried
out the calculations for the Co$^{2+}$ ion in the CoO$_{6}$
octahedral cluster using the XTLS8.3 program\cite{Tanaka94} with
parameter values typical for a Co$^{2+}$ system.\cite{parameters}
The spectra are dominated by the Co $2p$ core-hole spin-orbit
coupling which splits them roughly in two parts, namely the
$L_{3}$ ($h\nu \approx 779$ eV) and $L_{2}$ ($h\nu \approx 794$
eV) white lines regions. The curves in Fig. 1 reproduce the main
results of Tanaka and Jo,\cite{Tanaka91} and
De Groot.\cite{deGroot94} We have not separately shown the two
spectra for the doublet and quartet states originating from the
$\tilde{J}=5/2$  but combined them. The energy splitting between
these two set of states is small enough compared to the energy
difference between the ground-state and the center of the
$\tilde{J}=5/2$ states so that this splitting is of no real
importance if the states are to be populated by temperature.
These calculations show that the differences are so large that an
experimental resolution of order several 100 meV would be more
than sufficient to discriminate the various initial states
differing 100 meV or less in energy. Important is only the
temperature and the statistics with which the spectra have to be
measured.

Nevertheless, it is rather surprising that to the best of our
knowledge no experimental data have been reported in the
literature so far to confirm the validity of the calculations, i.e.
14-17 years after the appearance of Tanaka and Jo\cite{Tanaka91}
and De Groot\cite{deGroot94} studies. Indeed, it turned out to be
difficult to obtain reliable XAS data of CoO. Standard methods to
measure the XAS signal failed so far: the total electron yield
(TEY) method cannot be applied since CoO is strongly charging at
low temperatures, and the fluorescence yield (FY) method suffers
from too strong self-absorption effects,\cite{Pellegrin93} causing
severe distortions in the spectral line shapes. We therefore have
set out to prepare CoO in thin film form on metallic substrates,
thin enough as to avoid charging problems upon measurement using
the TEY method and yet thick enough to essentially retain the
properties of bulk CoO. The thin film must also be
polycrystalline, as to avoid complications caused by the
occurrence of (magnetic) linear dichroism
effects.\cite{Alders95,Alders98,Csiszar05}

The XAS measurements were performed at the Dragon beamline of the
NSRRC in Taiwan. The spectra were recorded using the total
electron yield method in a chamber with a base pressure of
3$\times$10$^{-10}$ mbar. The photon energy resolution at the Co
$L_{2,3}$ edges ($h\nu \approx 770-800$ eV) was set at 0.3 eV.
The actual polycrystalline CoO thin film was grown
\textit{in-situ} on a polycrystalline Ag sample by means of
molecular beam epitaxy (MBE), i.e. evaporating elemental and Co
from alumina crucibles in a pure oxygen atmosphere of $10^{-7}$
to $10^{-6}$ mbar. The base pressure of the MBE system is in the
low $10^{-10}$ mbar range. The thickness of the film was about 90
$\dot{A}$ as determined using a quartz-balance monitor, which in
turn was calibrated using the time period of the oscillation in
the intensity of reflection high energy electron diffraction
(RHEED) pattern of CoO thin films epitaxially grown on
Ag(001).\cite{Csiszar05,Csiszar05b} From that single crystal work
we also know that a 90 $\dot{A}$ CoO film has a N\'{e}el
temperature of about 290 K, i.e. essentially identical to the 291
K value determined for the bulk material.\cite{Singer56} The
polycrystalline CoO thin film has been measured at normal
incidence with a beam-spot size of about 1 mm$^2$. Homogeneity of
the thin film has been verified by measuring at several different
positions on the thin film resulting in negligible differences of
the spectra.

\begin{figure}
    \includegraphics[width=0.45\textwidth]{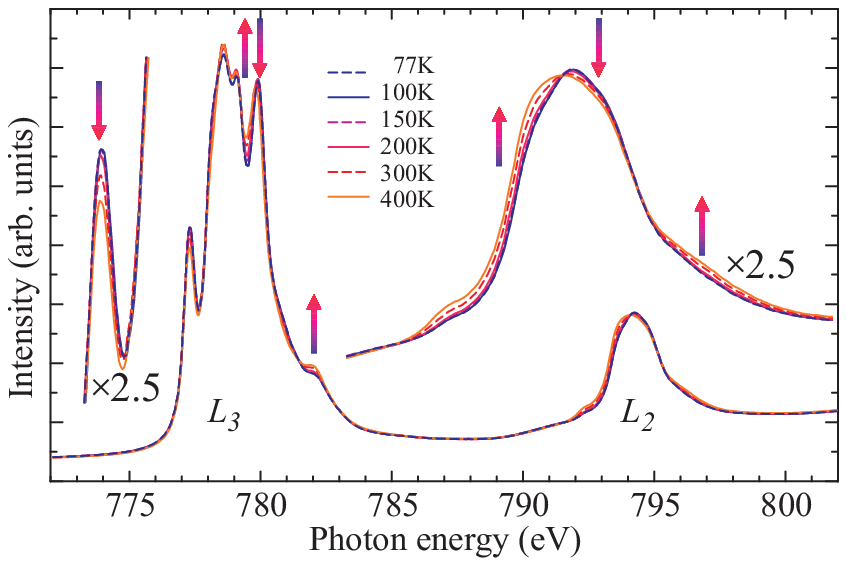}
    \caption{(color online) Experimental Co-$L_{2,3}$ XAS spectra of
    polycrystalline CoO thin film taken at various temperatures between 77 K
    and 400K.}
    \label{fig2}
\end{figure}

Fig. 2 shows the Co-$L_{2,3}$ XAS spectra of the polycrystalline
CoO thin film taken at various temperatures between 77 K and 400
K. The general line shape of the spectra shows the characteristic
features very similar to that of bulk CoO \cite{deGroot94},
verifying the good quality of our CoO film. The low-temperature
spectra have been measured before and after heating to 400 K
showing no detectable changes demonstrating the chemical
stability of the sample. Important is that high quality spectra
can be obtained for all temperatures, and this without the
slightest indication for a reduction of the intensity for the
lower temperature spectra, demonstrating that our thin film
approach is a successful resolution against the charging problems
often encountered when working with CoO.

\begin{figure}
    \includegraphics[width=0.45\textwidth]{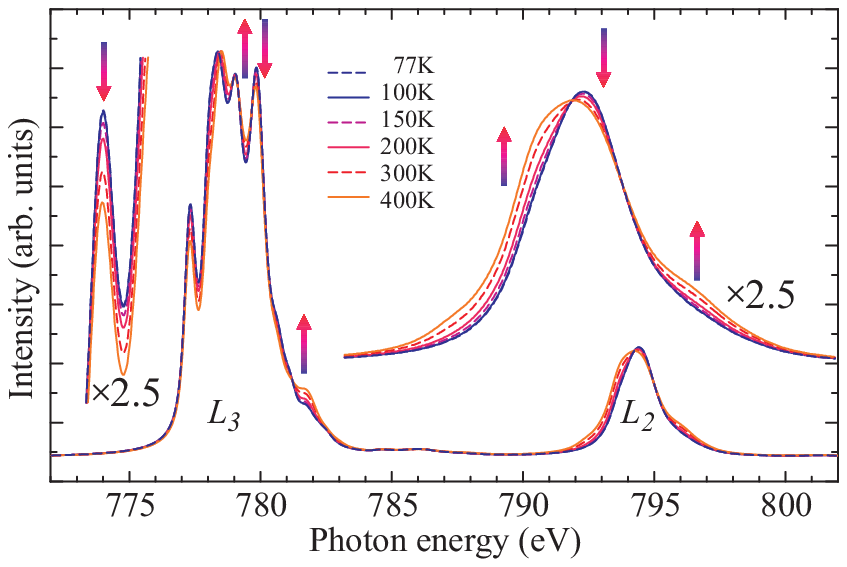}
    \caption{(color online) Theoretical Co-$L_{2,3}$ XAS spectra of
    CoO at several temperatures. Temperature dependence
    is due to population of excited states according to Bolzman statistics. Non-cubic distortions
    and magnetic correlations have been neglected.}
    \label{fig3}
\end{figure}

A closer look at the experimental spectra in Fig. 2 reveals that
there is a small but systematic change in the line shape with
temperature. A blow-up of the spectra in the $L_2$ edge range
makes this even clearer. To understand this temperature
dependence quantitatively, we simulated the spectra by summing up
the theoretical spectra starting from the different excited
states (see Fig. 1), weighted by their thermal population
according to the Boltzman statistics. These theoretical results
are plotted in Fig. 3, and one can observe a very good
agreement with the experiment. The predictions by Tanaka and
Jo\cite{Tanaka91} and De Groot\cite{deGroot94} are thus confirmed.

The calculations in Fig. 3 are done in cubic symmetry neglecting magnetic ordering. It is well known that CoO undergoes an anti-ferromagnetic ordering at 291 K. This phase transition is
accompanied by a tetragonal distortion and smaller distortions to
even lower symmetry. Tanaka and Jo \cite{Tanaka91} have shown
that the non-cubic distortions (and the exchange field) can mix
the low-energy states with different $t_{2g}$ orbital occupation,
split by spin-orbit coupling, thereby changing the lineshape of
the spectra. As the magnetic correlations and the tetragonal
distortion are temperature dependent this could also result in a
temperature dependence of the spectra. We will show below that the
inclusion of these effects will even improve the already very good
agreement between the temperature dependent theoretical and
experimental Co $L_{2,3}$ spectra as shown in Fig. 3.

\begin{figure*}
    \includegraphics[width=0.9\textwidth]{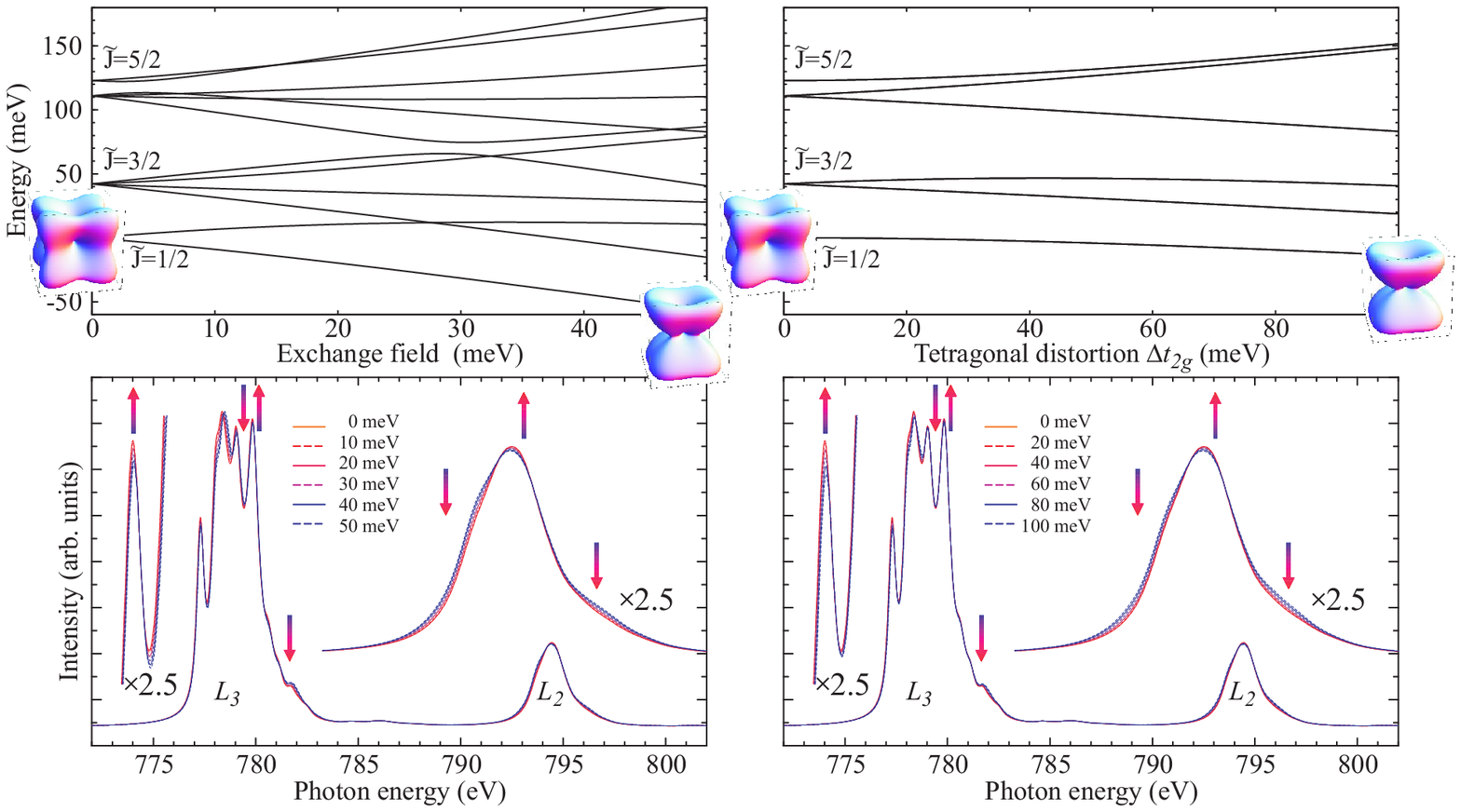}
    \caption{(color online) Top panels energy level diagram of the 12 lowest states of a CoO$_6^{10-}$
     cluster as a function of exchange field (left) and tetragonal crystal field distortion (right).
     Inset shows the ground-state $t_{2g}$ hole density.
     Bottom panels: Theoretical Co-$L_{2,3}$ XAS spectra of
     CoO as a function of exchange fields (left) and tetragonal crystal field distortion (right).}
    \label{fig4}
\end{figure*}

In the top panels of Fig. 4 we show the energy level diagram of
the lowest 12 eigen-states of a CoO$_6^{10-}$ cluster as a
function of exchange field (left) and tetragonal splitting
between the $d_{xy}$ and $d_{xz/yz}$ orbital (right). The inset
shows the $t_{2g}$ hole density. For cubic symmetry the
ground-state is a Kramer's-doublet, which has a cubic charge
density. For both an exchange field (in the $z$ direction) and a
tetragonal contraction of one Co-O bond, the $d_{xy}$ orbital
becomes more occupied and the hole resides in a complex linear
combination of the $d_{xz}$ and $d_{yz}$ orbital carrying an
orbital momentum parallel to the spin of that hole. The bottom
panels of Fig. 4 show the changes of spectra (at 0 K) as a
function of these non cubic distortions. One can clearly see
small changes in the spectra. The line-shape of the changes due to
quantum mechanical mixing of the excited states into the
ground-state are rather similar to the changes caused by thermal
population. This leads to a compensating effect: in going from
low to high temperatures, one has a reduction of the exchange
field and tetragonal splitting (they eventually disappear above
the N\'{e}el temperature), and thus changes in a direction
opposite as those caused by the thermal population effect. The
temperature dependence in the experimental spectrum is thus a
result of the thermal population of excited states as shown in Fig.
3, slightly reduced in its magnitude by quantum mechanical mixing
of the excited states into the ground-state at low temperatures as
shown in Fig. 4. Indeed, a detailed comparison between Figs. 2
and 3 reveal that the theoretical spectra show too much change
with temperature, and the inclusion of those quantum mechanical
mixing effects from Fig. 4 will improve the agreement even more.

\begin{figure}
    \includegraphics[width=0.5\textwidth]{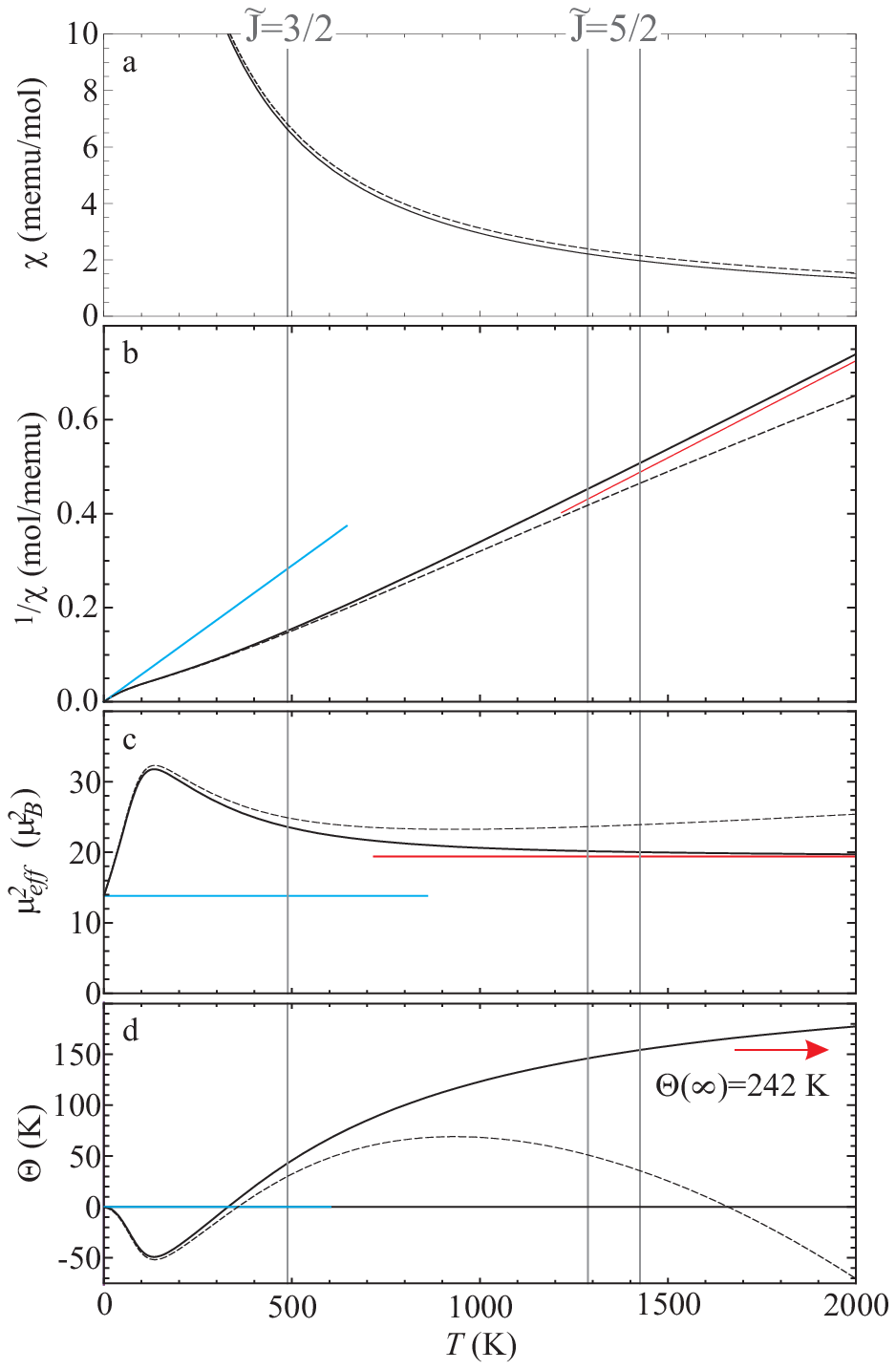}
    \caption{(color online) Calculated magnetic susceptibility and its derivatives for a single Co$^{2+}$ ion in cubic ligand field symmetry. Dotted lines without further correction, Straight lines Van Vleck (and Curie) moments induced by high energy (1 eV and above) excited states neglected. The vertical lines mark the energies of the excited states. The horizontal lines mark the low (blue) and high (red) temperature limit of the Van Vleck subtracted susceptibility. Panel a: Susceptibility. Panel b: Inverse susceptibility. Panel c: The (apparent) effective
             moment $\mu_{\textrm{eff}}^2$. Panel d: The (apparent) Weiss
             temperature $\Theta$ as defined in the text.}
    \label{fig5}
\end{figure}

The presence of the SOC has obviously important consequences for
the ground state and the magnetic properties of the Co ion. The
SOC lifts the orbital degeneracy and the ground-state (without
magnetic interactions) becomes a Kramers doublet with an
effective $\tilde{J}=1/2$. One should note that this Kramers
doublet is not Jahn-Teller active, that it has a cubic charge
density and does not show linear dichroism. This Kramers doublet has a magnetic moment, as revealed by the cluster calculations of $L_z=-0.56 \hbar$ and $S_z=-0.79 \hbar$ ($M_z=2.15 \mu_B$, $g=4.29$, and $\mu_{\textrm{eff}}^2=3.7^2 \mu_B^2$). The induced Van Vleck moment is $0.018 \mu_B$ per Tesla. Exchange fields and crystal distortions mix in excited states, thereby altering the expectation value of the magnetic moment. At the same time these distortions induce a non-cubic local charge density and create linear dichroism. In order to calculate the low temperature ordered moments we need to make an estimate for the exchange field and tetragonal distortion. CoO has a 1.2\% tetragonal contraction at low T. \cite{Roth58} If we interpolate the values for the crystal-field and the exchange field as found for CoO thin films grown on MnO and Ag \cite{Csiszar05} we find an exchange field of 12.6 meV and a tetragonal distortion of 25 meV between the $d_{xy}$ and the $d_{xz/yz}$ orbital. With these values we find for the ground-state that the total magnetic moments are $S_z=-1.14 \hbar$ and $L_z=-1.15 \hbar$ ($M_z=3.44 \mu_B$).

The SOC induced splitting of the low lying states has also
important consequences for the temperature dependence of the
magnetic susceptibility, in an analogous manner as it causes the
temperature dependence of the XAS spectra. We have calculated the magnetic susceptibility $\chi=\lim_{H\rightarrow0} \partial M/ \partial H$ (which is equal to $M/H$ for $H$ small enough), for a CoO$_6^{10-}$ cluster. In Fig. 5a we show the the susceptibility calculated for the entire cluster including all Van Vleck moments (dotted line) and the susceptibility calculated for the lowest 12 eigenstates (straight line). For the second calculation we have thus neglected the Van Vleck (and Curie) susceptibility induced by the higher lying states, such as found around 1 eV involving a $t_{2g}$-$e_{g}$ orbital excitation. This roughly corresponds to experimentally subtracting a constant from the measured susceptibility such that the high temperature limit goes to zero. Panel b of Fig. 5 shows $1/\chi$ against the
temperature. One can observe that the curve does not follow a
linear behavior in a large temperature regime. There are two limiting cases of interest. For $k_{B} T$ much smaller then the energy of the first excited state one should expect normal Curie-Weiss behaviour. This asymptote is indicated in Fig. 5b by a straight line on the left (blue). The other limit would be when $k_{B} T$ is much larger then the highest energy of the excited states split by SOC. In this limit one would again expect to retrieve a 're-normalized' Curie-Weiss like behavior. This limit is indicated in Fig. 5b by a straight line on the right hand side (red). One can see that for temperatures starting roughly at 600 K $1/\chi$ shows linear behaviour and the high temperature limit is reasonably achieved. One might naively
think that the high temperature region can be used to extract
magnetic quantum numbers as it is often done when a
Curie-Weiss-like behavior is observed, but then one must also
realize that those numbers are not representative for the ground
state.

To illustrate this more clearly, we will discuss both limits in more detail. We start with the low temperature limit. In Fig. 5c we plot $\mu_{\textrm{eff}}^2$, the (apparent) effective magnetic moment squared ($\mu_{\textrm{eff}}^2$ is defined here as $3k_B$ divided by the temperature derivative of $1/\chi(T)$). In the low temperature limit $\mu_{\textrm{eff}}^2$ does not become a constant as one should expect for Curie-Weiss behaviour as we did not subtract the Van Vleck moments for the ground-state doublet. (The straight line is calculated for a subtraction of the Van Vleck moments for the lowest 12 states, which still allows the ground-state doublet to have a positive Van Vleck moment and the excited states a negative Van Vleck moment such that the sum of the lowest twelve states equals zero.) The linear behaviour of $\mu_{\textrm{eff}}^2$ for small $T$ is easy to understand if one realizes that $\chi(T)=\mu_{1}^2/(3k_B T)+\chi_{VV}$, with $\chi_{VV}$ the constant Van Vleck susceptibility and $\mu_{1}$ the effective moment for the ground state. Then $1/\chi(T)=(3k_B T)/\mu_{1}^2 - (3^2k_B^2 T^2 \chi_{VV})/\mu_{1}^4 + O(T^3)$ and $\mu_{\textrm{eff}}^2 \equiv 3k_B / \partial_T (1/\chi)=\mu_{1}^2 + 6 \chi_{VV} k_B T + O(T^2)$.

The high temperature limit can be understood if one neglects Van Vleck contributions for all magnetic moments. One then finds that the average effective moment squared is just the average of the effective moments squared of each excited stated, weighted by their multiplicity: $\overline{\mu_N^2}={\sum_{i=1}^N\nu_i\mu_i^2}/{\sum_{i=1}^N\nu_i}$. Here $\overline{\mu_N^2}$ is the high temperature limit for the average effective moment squared of the first $N$ non-degenerate sets of eigenstates, $\mu_i^2$ the effective moment squared of the $i$-th non-degenerate set of eigenstates and $\nu_i$ the degeneracy of the $i$-th set of eigenstates. For a cluster calculation of Co$^{2+}$ one finds a high temperature effective moment of $\mu_{\textrm{eff}}^2 = 4.4^2 \mu_B^2$ while the ground state value is $ 3.7^2 \mu_B^2$. One does not only find a deviation from the ground-state effective moment but the high temperature limit also induces an apparent Weiss temperature. The Weiss temperature $\Theta$ is defined here as the intercept of the tangent to the $1/\chi(T)$ curve with the abscissa. This can be calculated by making a series expansion of $1/\chi$ in $1/T$. $\Theta_N={1}/{k_B}\left({\sum_{i=1}^N\nu_i\Delta_i}/{\sum_{i=1}^N\nu_i}-{\sum_{i=1}^N\nu_i\Delta_i\mu_i^2}/{\sum_{i=1}^N\nu_i\mu_i^2}\right)$, with $\Delta_i$ the energy of the $i$-th set of eigenstates. In panel d of Fig. 5 we plot the apparent Weiss temperature as a function of T. One can see a strong temperature dependence and the Weiss temperature slowly saturates at 242 K. This apparent Weiss temperature has nothing to do with magnetic correlations since they were not included in this single ion calculation. Instead, the
strongly varying (apparent) Weiss temperature merely reflects the
fact that there are excited states with different magnetic
quantum numbers thermally populated.

Interestingly one finds that the high temperature limit for the (apparent) effective moment squared is reached at temperatures much lower then the energy of the highest excited state. We have added vertical bars at the energies of the excited states in Fig. 5 and one can see that $\mu_{\textrm{eff}}^2$ becomes reasonable constant at a temperature that is about half the energy of the highest excited state. Not only does one find that the high temperature regime for $\mu_{\textrm{eff}}^2$ starts at rather low temperatures, one also finds that the low temperature region, where the ground-state Curie-Weiss and Van Vleck moments can be obtained, extends only up to 50K, roughly a tenth of the energy of the first excited state. This is particular disturbing as one naively would expect that excited states would only distort the susceptibility at temperatures comparable to the energy of the excited state.

In the presence of magnetic correlations, such as in the real CoO
material, those thermal population effects become very much masked
or hidden. One then easily interprets the high temperature
Curie-Weiss limit as a region in which one is free from magnetic
correlations and yet still can capture the ground state local
quantum numbers, and one also easily thinks that the measured
Weiss temperature is only due to the magnetic correlations. The
application of the Curie-Weiss law obviously leads to errors in
case SOC is active.

The effect of SOC excited states on the magnetic susceptibility are well recognized in systems with an open $4f$ shell. \cite{4fexamples} For $3d$ systems with an open $t_{2g}$ shell in nearly cubic symmetry these effects should also be important. In that respect high-spin Fe$^{2+}$, Co$^{3+}$ and Co$^{2+}$ are good candidates to show such effects.

To conclude, we have been able to accurately measure the
temperature dependence of the Co $L_{2,3}$ x-ray absorption
spectra of CoO. The use of polycrystalline thin films of CoO grown
on a Ag substrate allowed us to avoid charging problems usually
encountered in electron spectroscopic studies on CoO. The changes
in spectra as a function of temperature can be well explained in
terms of the thermal population of closely lying excited states,
originating from degenerate $t_{2g}$ levels lifted by the
spin-orbit coupling. The existence of these excited states also
lead to a non-trivial temperature dependence for the magnetic
susceptibility, e.g. one cannot apply the Curie-Weiss law to
extract the relevant \textit{ground} state quantum numbers from
the high temperature region.

\end{document}